# Multiferroicity in the two-dimensional limit in hexagonal $LuFeO_3$ films

Huilin Lai[1,2], Junyu Tan[1,2], Jinfeng Zhai[1,2], Yang Shi[1,2], Lili Feng[1,2], Huanyu Zhang[1,2], Chuanrui Huo[3], Chuhang Liu[4], Lijun Wu[4], Lifeng Yin[1,2,5,6,7], Hangwen Guo[1,2,5], Jun Chen[3], Xiaoshan Xu[8], Jun Zhao[1,2,5,6], Yimei Zhu[4], Shiqing Deng[3*], Wenbin Wang[1,2,5*] and Jian Shen[1,2,5,6,7*]

[1]State Key Laboratory of Surface Physics and Institute for Nanoelectronic Devices and Quantum Computing, Fudan University, Shanghai 200433, China.

[2]Department of Physics, Fudan University, Shanghai 200433, China.

[3]Beijing Advanced Innovation Center for Materials Genome Engineering, University of Science and Technology Beijing, Beijing, 100083, P.R. China.

[4]Condensed Matter Physics and Materials Science Department, Brookhaven National Laboratory, Upton, NY 11973, USA.

[5]Hefei National Laboratory, Hefei 230088, China.

[6]Shanghai Research Center for Quantum Sciences, Shanghai 201315, China.

[7]Zhangjiang Fudan International Innovation Center, Fudan University, Shanghai 201210, China.

[8]Department of Physics and Astronomy, University of Nebraska, Lincoln, Nebraska 68588, USA.

*Corresponding author. Email: sqdeng@ustb.edu.cn; wangwb@fudan.edu.cn; shenj5494@fudan.edu.cn

**Abstract:** Multiferroic oxides, which host coexisting ferroelectric and magnetic orders, are central to understanding correlated quantum phenomena. Yet, as thickness approaches the two-dimensional (2D) limit, both ferroelectricity and magnetism are generally expected to be strongly suppressed by depolarization fields and finite-size effects, respectively. Here, we show that hexagonal $LuFeO_3$ ($h$-$LuFeO_3$) retains multiferroic behavior down to a thickness of only one and a half unit cells. The polar structural distortion remains robust at room temperature and is comparable to that of the bulk, while robust magnetic order persists at low temperatures. We further find that the magnetic response can be reproducibly modulated through ferroelectric domain-state control. Structural analysis indicates that the $K_3$ distortion remains stable down to the two-dimensional limit, providing the basis for the survival of improper ferroelectricity in this regime. These results establish $h$-$LuFeO_3$ as an oxide platform in which ferroelectricity and magnetic order coexist at the atomic-scale limit, and provide insight into how coupled ferroic behavior can persist in ultrathin complex oxides.

The exploration of multiferroic materials remains a central focus in condensed matter physics because the coexistence of electric and magnetic orders provides a valuable platform for investigating correlated phenomena and emergent interactions[1-4]. A major frontier in this field is to determine whether multiferroicity can survive when dimensionality is reduced to the atomic limit. In ultrathin systems, ferroelectricity is typically suppressed by depolarization fields[5], while magnetic order is destabilized by finite-size effects[6,7]. Whether both ferroic orders can coexist and whether their coupling can remain operative at the two-dimensional (2D) limit therefore remains a fundamental open question. Although multiferroicity has also been explored in two-dimensional van der Waals materials, issues of environmental stability, reproducibility, and device compatibility remain important challenges for practical implementation[8-12]. In contrast, oxide multiferroics provide a complementary platform with robust chemical stability, structural tunability, and compatibility with epitaxial interface engineering[13-15]. These features make oxide systems particularly attractive for testing whether multiferroic behavior can persist at the atomic-scale limit.

Recent progress has demonstrated that ultrathin oxide films can retain non-centrosymmetric structures supporting ferroelectricity, in some cases even approaching the monolayer limit[16-18]. However, direct experimental realization of multiferroicity at the atomic-scale limit has remained elusive. Most prior studies have either confirmed ferroelectric order alone or required thicker films to detect magnetic responses[19-24]. As a result, whether ferroelectricity and magnetic order can survive together in an oxide system at the true 2D limit remains unresolved.

Hexagonal $LuFeO_3$ ($h$-$LuFeO_3$) is a particularly attractive platform for addressing this question. As an improper ferroelectric, its polar state is tied to the nonpolar $K_3$ structural distortion rather than driven by a conventional soft polar mode. This makes the polar distortion intrinsically less sensitive to depolarization effects than in proper ferroelectrics[25]. In real ultrathin films, however, substrate-induced interfacial clamping can still suppress the $K_3$ distortion and destabilize the polar phase[20,24]. A key question, therefore, is whether this improper-ferroelectric framework can preserve the polar distortion and, at the same time, sustain magnetic order so that multiferroic behavior survives at the true 2D limit.

In this work, we show that $h$-$LuFeO_3$ retains robust ferroelectricity together with persistent magnetic order down to a thickness of one and a half unit cells (1.5 UC). Ferroelectric measurements reveal switchable polarization in the 1.5 UC film, supported by the persistence of polar structural distortion observed by atomic-scale structural characterization. By leveraging the proximity effect in Pt/$h$-$LuFeO_3$ heterostructures, we further detect magnetic ordering through anisotropic magnetoresistance (AMR) and anomalous Hall effect (AHE) measurements. More importantly, we find that the magnetic response can be reproducibly modulated through control of the ferroelectric domain state. These results establish h-$LuFeO_3$ as an oxide platform in which multiferroic behavior persists at the atomic-scale limit.

$H$-$LuFeO_3$ exhibits a layered structure with alternating Fe-O and Lu-O planes oriented along the [001] direction. Each unit cell consists of two Fe-O layers and two Lu-O layers. The Fe atoms are arranged in a planar triangular lattice, forming $FeO_5$ trigonal bipyramids[26], as illustrated in the inset in Fig. **1(a)**. A cooperative rotation and tilting of these bipyramids, corresponding to the nonpolar $K_3$ distortion, drives the characteristic trimerized structure of the hexagonal phase. Through its coupling to the polar $\Gamma_2^-$ mode, this distortion gives rise to the polar structure and geometric improper ferroelectricity in $h$-$LuFeO_3$ [24,27-31]. In addition, $h$-$LuFeO_3$ exhibits canted antiferromagnetic order, in which the Dzyaloshinskii-Moriya interaction produces a weak net spin component along the c axis [26,32-35].

A series of thin films with different thicknesses up to 22 UC (25.7nm) were synthesized using the pulsed laser deposition (PLD) technique. Initially, a three-layer $Lu_2O_3$ buffer was epitaxially deposited on a YSZ (111) substrate, followed by the deposition of $h$-$LuFeO_3$ film. The $Lu_2O_3$ buffer reengineers the interface and helps alleviate substrate-induced clamping, thereby preserving the $K_3$ related polar distortion and improving the structural stability of $h$-$LuFeO_3$ even in the ultrathin regime. During thin film deposition, the film growth process was monitored in real time using reflection high-energy electron diffraction (RHEED), with one RHEED oscillation corresponding to the completion of half a unit cell of $h$-$LuFeO_3$[36]. X-ray reflectivity (XRR) measurements were employed to calibrate the film thickness[36], confirming the one-to-one correspondence between the RHEED oscillation period and the sample thickness.

Atomic-scale scanning transmission electron microscopy (STEM) was employed to investigate the film structure. Fig. **1(a)** displays a typical high-angle annular dark-field (HAADF) image of the $h$-$LuFeO_3$ film, where both upward and downward polarization domains are clearly observed. The accompanying schematic on the right illustrates the thin film structure, which consists of a 1.5 UC $h$-$LuFeO_3$ layer and a three-atomic-layer $Lu_2O_3$ buffer layer grown on a YSZ [111] substrate. Fig. **1(b)** presents electron energy loss spectroscopy (EELS) mappings of Lu $M_{4,5}$, Fe $L_{2,3}$, Zr $L_{2,3}$, Pt $M_{4,5}$ and O $K$ edges, which reveal distinct compositional profiles and confirm the structural integrity of the film. Fig. **1(c)** and **1(d)** show enlarged views that highlight the upward and downward polarization domains, respectively. In $h$-$LuFeO_3$, the paraelectric-to-ferroelectric transition involves in-plane unit cell tripling for the ferroelectric phase ($P6_3cm$ phase) relative to the paraelectric phase ($P6_3/mmc$), characterized by an "up–up–down" or "down–down–up" displacement of Lu atoms along [001] [37,38]. Such buckling of the Lu layer is evident in the 1.5 UC sample, indicating the retention of the polar structure of $h$-$LuFeO_3$. The Lu displacements were further quantified from the HAADF images, yielding an out-of-plane displacement magnitudes of ~44 pm. This value is comparable to that of bulk (~41 pm for $Lu_{0.5}Sc_{0.5}FeO_3$ single crystal)[37], indicating that the characteristic polar structural distortion remains essentially preserved. Although the Lu displacement is not a direct quantitative measure of polarization, its comparable magnitude strongly suggests that room-temperature ferroelectricity remains robust in the 1.5 UC film, with no obvious suppression of the polar distortion by depolarization effects.

Fig. **2(a)** shows the RHEED pattern for the 1.5 UC $h$-$LuFeO_3$, where additional streaks appear at the 1/3 and 2/3 positions indicating the in-plane unit cell tripling ($P6_3cm$ phase) relative to the paraelectric phase ($P6_3/mmc$) (Fig. **2(b)**). Together with the STEM observations, the presence of a ferroelectric polar structure is confirmed in the 1.5 UC $h$-$LuFeO_3$ film. These results suggest the absence of critical thickness in improper ferroelectric films[25].

To verify the RT ferroelectricity and its switchability, we performed piezo-response force microscopy (PFM) measurements on the 1.5 UC $h$-$LuFeO_3$ film[36]. Fig. **2(c)** and **2(d)** show PFM phase and amplitude images scanned under +3V and -3V tip bias in different regions over a 3×3 $\mu m^2$ area, respectively. Stable remanent polarization with clear bipolar contrast is observed, indicating reversible polarization states of the film. Local PFM switching spectroscopy demonstrated robust ferroelectric switching behavior in the 1.5 UC $h$-$LuFeO_3$. As shown in Fig.

**2(e)** and **2(f)**, with switching voltages around 0.4 V, a well-defined 180° phase hysteresis loop and butterfly-shaped amplitude ($d_{33}$) loops are observed, suggesting stable polarization switching. Time-dependent PFM imaging was also conducted to evaluate the retention properties of the film. The results show that stable polarization patterns persist over 3.5 hours, confirming the exceptional retention and long-term stability of the 1.5 UC $h$-$LuFeO_3$[36]. Detailed analysis show that the temporal evolution of the ferroelectric polarization follows a power-law decay[18,39,40]: $P(t) = P_0 \cdot t^{-\alpha}$ where $P$(t) is the polarization strength at time $t$, $P_0$ is the initial polarization, and α is the power-law exponent. The extracted exponent, $\alpha$=0.028, is remarkably small, indicating that the polarization decays extremely slowly over time. These results highlight the robustness of the ferroelectricity in the 1.5 UC $h$-$LuFeO_3$ film, underscoring their potential for long-term stability in applications.

The bulk $h$-$LuFeO_3$ is a spin-frustrated system, exhibiting weak net magnetic moments at 130 K due to the canted antiferromagnetic ordering caused by DM interactions[26]. Considering the fact that the already weak magnetization becomes even more difficult to detect in ultrathin limit, we use magnetic proximity effect induced AMR to probe the magnetic properties of the 1.5 UC $h$-$LuFeO_3$ film. Specifically, Pt layers were deposited on $h$-$LuFeO_3$ films. Pt conduction electrons align with the adjacent magnetic moments of $h$-$LuFeO_3$, inducing spin polarization that sensitively reflects the films' magnetic ordering[41,42]. For our experiments, a 4.5 nm Pt layer was applied, and the samples were fabricated into Hall bar devices (Fig. **3(a)**). To validate our method, we demonstrate that the AMR response of a thick film (22 UC) exhibits bulk-like magnetic transitions, consistent with superconducting quantum interference device (SQUID) measurements. As shown in Fig. **3(c)**, the AMR signal appears below 130 K, matching the transition ($T_N$= 130 K) observed in zero-field-cooled (ZFC) and field-cooled (FC) SQUID measurements (Fig. **3(b)**). These results, in agreement with previous reports on thicker films[26], confirm that AMR in Pt, driven by interfacial proximity effects, is a reliable technique for probing the magnetic properties of the ultrathin $h$-$LuFeO_3$ films.

Figures **3(d)–(f)** show the temperature dependence of the AMR signal for samples with thicknesses of 9 UC, 4 UC, and 1.5 UC, respectively. The AMR signal is observed to vanish above the magnetic transition temperature $T_N$. Fig. **3(g)** depicts the $T_N$ for samples of varying thicknesses. It is evident that as the film thickness decreases, $T_N$ monotonically decreases from the bulk value of 130 K in the 22 UC sample to 25 K in the 1.5 UC film. Furthermore, at a fixed temperature of 10 K, the amplitude of the AMR signal also decreases with reducing thickness (Fig. **3(g)**, inset). This behavior aligns with the universal scaling law commonly observed in magnetic thin films (Fig. **3(h)**). Specifically, the shift in $T_N$ follows a power-law dependence for thicker films: $t(N) = 1 - T_N(N)/T_N(\infty) \sim N^{-\lambda}$, where N is the number of monolayers and λ is the scaling exponent. As the film thickness approaches the ultrathin limit, the behavior transits to a linear dependence of $t(N) \sim N$. These results are in good agreement with theoretical predictions for magnetic thin films[6,7,43,44]. Importantly, even in the 2D limit, relatively strong magnetism is preserved in the 1.5 UC sample, highlighting the remarkable 2D magnetic properties of the $h$-$LuFeO_3$ system.

To further confirm that the observed AMR signals are intrinsic to $h$-$LuFeO_3$, we introduced a 2 nm Cu spacer layer between the Pt and $h$-$LuFeO_3$ layers to block magnetic proximity effects[45]. In this configuration, the AMR signals vanish across the entire temperature range[36], confirming that the Pt-AMR measurements accurately reflect the inherent magnetic properties of $h$-$LuFeO_3$.

To probe the coupling between ferroelectric and magnetic degrees of freedom, we applied an external electric field to align the ferroelectric polarization direction of the sample and then examined the corresponding magnetic responses. Specifically, an electric field of 1.5 V/100 nm was applied to polarize the film[36]. After poling, the electric field was removed, and all subsequent magnetic measurements were carried out in zero field. As shown in Fig. **4(a)**, AMR measurements at 10 K revealed a significant enhancement in the signal for the electrically poled 1.5-UC h-$LuFeO_3$ film, with polarization aligned along +Z, compared to the initial unpoled state with spontaneous polarization. This enhancement indicates an increase in the magnetic moment along the $c$-axis. To further investigate this effect, we also measured the AHE in the Pt layer at 10 K, which similarly reflects the magnetism of $h$-$LuFeO_3$ through the magnetic proximity effect at the interface. The AHE signal is likewise enhanced after electric poling, indicating an increased net magnetic moment in the poled state (Fig. **4(b)**).

Next, we applied an electric field of equal magnitude but in the reverse direction (-Z) to the sample. This reversal led to a complete switching of the ferroelectric domains, polarizing the system along the -Z direction. Under these conditions, the AHE signal remains essentially the same as that obtained after +Z poling. This indicates that reversing the ferroelectric polarization direction does not reverse the sign of the magnetic response. In contrast, when the poled sample is allowed to relax for an extended time, the ferroelectric state evolves away from the saturated single-domain configuration toward a multidomain state, as indicated by the ferroelectric retention measurements. Correspondingly, the AHE signal is significantly reduced after such relaxation (Fig. **4(b)**). These results establish a direct correlation between the ferroelectric domain state and the magnetic response: the magnetic signal is maximized in the saturated single-domain state and suppressed in the relaxed multidomain state.

This ferroelectric domain–dependent modulation of the magnetic response highlights a distinctive form of intrinsic magnetoelectric coupling in $h$-$LuFeO_3$[46-49]. In this system, both ferroelectricity and magnetism originate from the same underlying non-centrosymmetric $K_3$ distortion. The $K_3$ distortion drives the trimerization of the $FeO_5$ bipyramids and the buckling of the Lu layers, thereby stabilizing the improper ferroelectric state. Through spin-lattice coupling, including single-ion anisotropy and the Dzyaloshinskii-Moriya interaction, the same lattice distortion also enables spin canting and a weak net moment along the c axis. Because both order parameters share this common structural origin, the magnetic response remains coupled to the ferroelectric domain configuration even in the ultrathin limit.

Beyond this shared structural origin, the topological nature of the ferroelectric domain structure provides a unique pathway for magnetoelectric coupling via domain walls[50-55] (Fig. **4(c)**). In hexagonal improper multiferroics, the phase $\Phi$ of the $K_3$ mode determines the structural trimerization, while the spin orientation $\psi$ is constrained by magnetic anisotropy as $M = -M_s\cos(\Phi - \psi)$, where $M_s$ is the saturated magnetization. Within a single ferroelectric domain, $\psi$ is locked to $\Phi$ with a phase difference of $n\pi$[33,52], giving the maximum weak ferromagnetic moment. Across a ferroelectric domain wall, however, $\Phi$ rotates by $\pi/3$, and $\psi$ is expected to evolve accordingly. This leads to a local mismatch between $\Phi$ and $\psi$ ($\Phi - \psi \neq n\pi$), thereby suppressing the weak ferromagnetic canting (Fig. **4(d)**), manifested as clamped antiferromagnetic walls. When the density of such walls increases, the overall magnetic response is correspondingly suppressed.

Our observations are fully consistent with this picture. The +Z and -Z saturated states exhibit the same AHE magnitude, showing that the ferroelectric polarization direction does not directly determine the sign of the weak ferromagnetic moment. By contrast, the multidomain state gives a substantially smaller AHE signal, consistent with a larger population of clamped antiferromagnetic walls and a reduced net moment. The reproducibility of this behavior across multiple samples indicates that it is not a stochastic or extrinsic effect, but rather a consequence of the intrinsic domain topology of *h*-$LuFeO_3$. We therefore interpret the observed behavior not as bulk magnetoelectric coupling effect, but as domain wall-mediated magnetoelectric coupling effect, which can be topologically governed, structurally encoded, and electrically tunable modulation of magnetic response through ferroelectric domain-state control.

In summary, we have shown that 1.5-UC *h*-$LuFeO_3$ retains robust ferroelectricity together with persistent magnetic order at the two-dimensional limit. The polar structural distortion remains comparable to that of the bulk, and the magnetic response can be reproducibly modulated through control of the ferroelectric domain state. These results indicate that the $K_3$ driven improper-ferroelectric framework, together with interface engineering, enables multiferroic behavior to persist despite the strong suppression effects usually encountered in ultrathin films. The films also remain stable after prolonged air exposure, underscoring the environmental robustness of this oxide system. Our findings establish *h*-$LuFeO_3$ as a model platform for exploring low-dimensional multiferroics and provide guidance for designing oxide-based spintronic and nanoelectronic devices at the atomic-scale limit.

**Data availability**

The data that support the findings of this study are available within the article and Supplementary Information.

**Acknowledgments**
This work was supported by the National Key Research Program of China (2022YFA1403300, 2020YFA0309100), the Innovation Program for Quantum Science and Technology (2024ZD0300103), the National Natural Science Foundation of China (12074071, 22375015), Shanghai Municipal Science and Technology Major Project (2019SHZDZX01), and Xiaomi Young Scholars Fund. The work at Brookhaven National Laboratory was supported by the Materials Science and Engineering Divisions, U.S. DOE-BES under Contract No. DE-SC0012704.

**Author contributions**

J.S. and W.W. conceived and designed the experiments. H.L. and W.W. fabricated the materials and prepared all devices. S.D., Y.Z., J.C., C.H., C.L. and L.W. performed characterization and analysis of microphysical properties. H.L. and J.T. performed the structural characterization. H.L., L.F., Y.S. and H.Z. performed the ferroelectric properties measurement. H.L. and J.-F.Z. performed the magnetic properties measurement. H.L. and W.W. conducted the magnetoelectric coupling measurement. W.W., H.L., J.S., S.D., X.X., J.Z., H.G. and L.Y. analyzed the data. W.W., J.S. and S.D. wrote the original manuscript. WW, JS, SD, HL, XX, JZ, HG and LY revised and edited the manuscript.

**Competing interests**

The authors declare no competing interests.

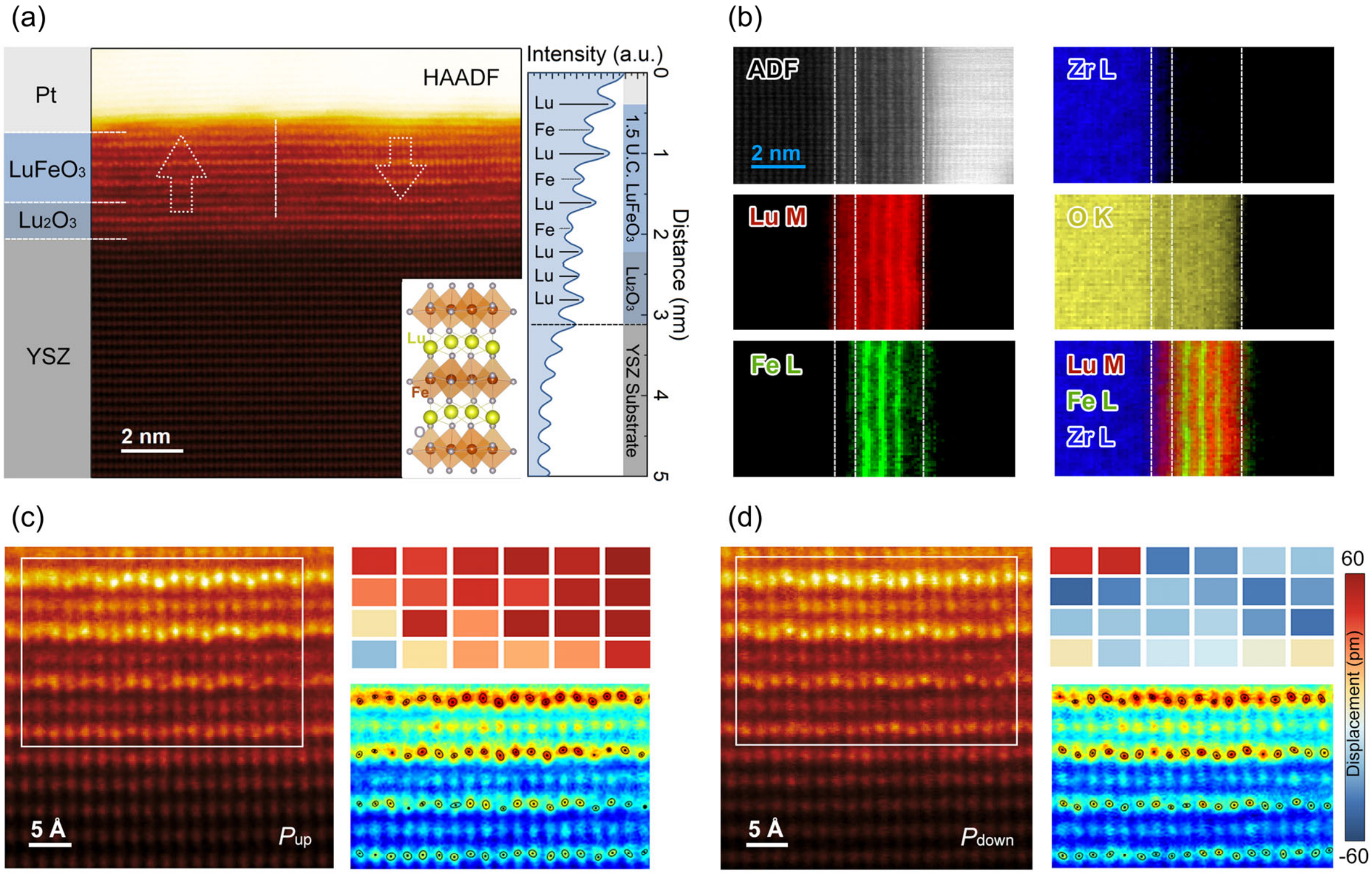

**Fig. 1.** (**a**) HAADF-STEM image revealing the atomic-scale structure of the *h*-$LuFeO_3$ thin film, where both the upward and downward polarization domains are observed. The left schematic illustrates the thin film structure, consisting of a 1.5-unit-cell-thick $LuFeO_3$ layer and a three-atomic-layer $Lu_2O_3$ buffer layer on the YSZ [111] substrate. Inset shows the crystal model of *h*-$LuFeO_3$. Right panel displays the integrated intensity profile obtained from the blue-shaded area in the HAADF-STEM image. (**b**) EELS elemental mapping for the Lu $M_{4,5}$, Fe $L_{2,3}$, Zr $L_{2,3}$, Pt $M_{4,5}$, and O $K$ edges. Note that due to the overlapping of Zr $L_{2,3}$ and Pt $M_{4,5}$, their signals are combined into a single map, without distinguishing them. (**c**, **d**) Representative enlarged views highlighting the upward and downward polarization domains. Right panels display the corresponding maps of fitted atomic positions (lower panel) and calculated out-of-plane Lu-ion displacement (upper panel) derived from the white rectangle-framed areas indicated in the left HAADF images.

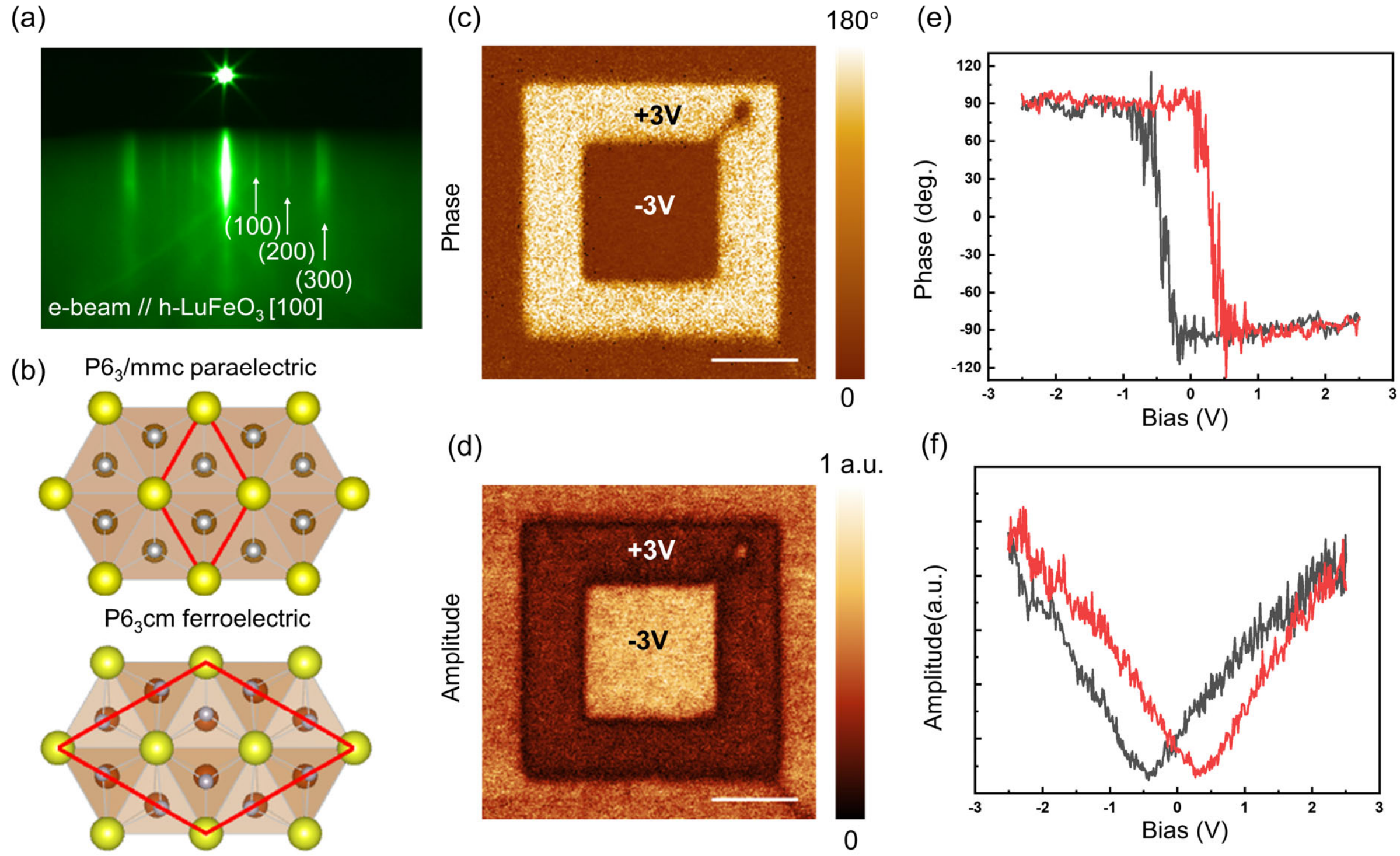

**Fig. 2.** (**a**) RHEED pattern of the 1.5 UC *h*-$LuFeO_3$, exhibiting additional streaks at the 1/3 and 2/3 positions. (**b**) Illustration of the in-plane unit cell tripling in the ferroelectric phase ($P6_3cm$) compared to the paraelectric phase ($P6_3/mmc$). The paraelectric phase exhibits a simple hexagonal structure, while the ferroelectric phase undergoes a $\sqrt{3} * \sqrt{3}$ reconstruction in the a-b plane due to structural distortions, as indicated by the red lines, resulting in a tripled unit cell. (**c**, **d**) PFM phase and amplitude images of ultrathin 1.5 UC *h*-$LuFeO_3$, obtained after sequential application of tip biases of +3V and -3V over a 5×5 $\mu m^2$ area, the scale bar is 1 $\mu m$. (**e**) Local PFM spectroscopy reveals a robust 180° phase hysteresis. (**f**) Butterfly-shaped amplitude ($d_{33}$) loops with switching voltages around 0.4 V.

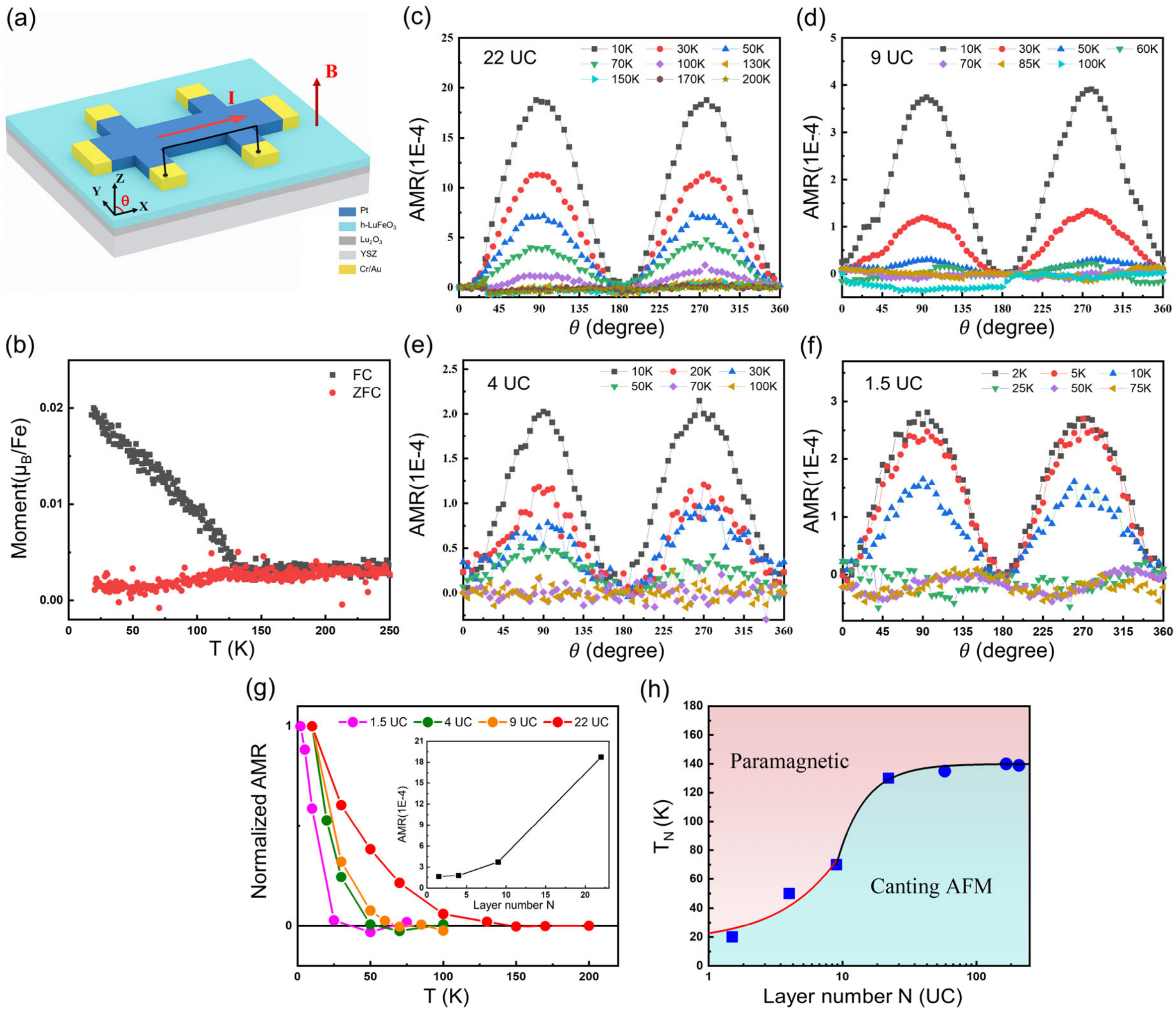

**Fig. 3.** (**a**) Schematic of the Hall bar device used for magnetic characterization. (**b**) Temperature-dependent ZFC and FC magnetization curves from SQUID measurements on the 22 UC samples, confirming a magnetic transition at 130 K. (**c-f**) Temperature-dependent AMR measurements performed on h-$LuFeO_3$ films with thicknesses of 22 UC, 9 UC, 4 UC, and 1.5 UC. (**g**) The AMR signal disappeared at approximately 130 K, 70 K, 50 K, and 25 K, respectively, corresponding to the magnetic transition temperatures ($T_N$) for each thickness. Insert: AMR signal strength at a constant measurement temperature of 10 K, showing a progressive reduction with decreasing sample thickness. (**h**) Phase diagram of *h*-$LuFeO_3$ magnetic transition temperature as a function of layer number. Square data points represent results from this work, while circular data points are taken from previous literature[32]. The black line represents a fit to the magnetic transition temperature ($T_N$) for samples with 9 UC or more ($N \geq 9$) using the finite-size scaling formula: $t(N) = 1 - T_N(N)/T_N(\infty) \sim N^{-\lambda},\ \lambda = 2.067$. The red line is a linear fit to $T_N$ for samples with 9 UC or fewer. The blue region corresponds to the canted antiferromagnetic phase, while the pink region represents the paramagnetic phase.

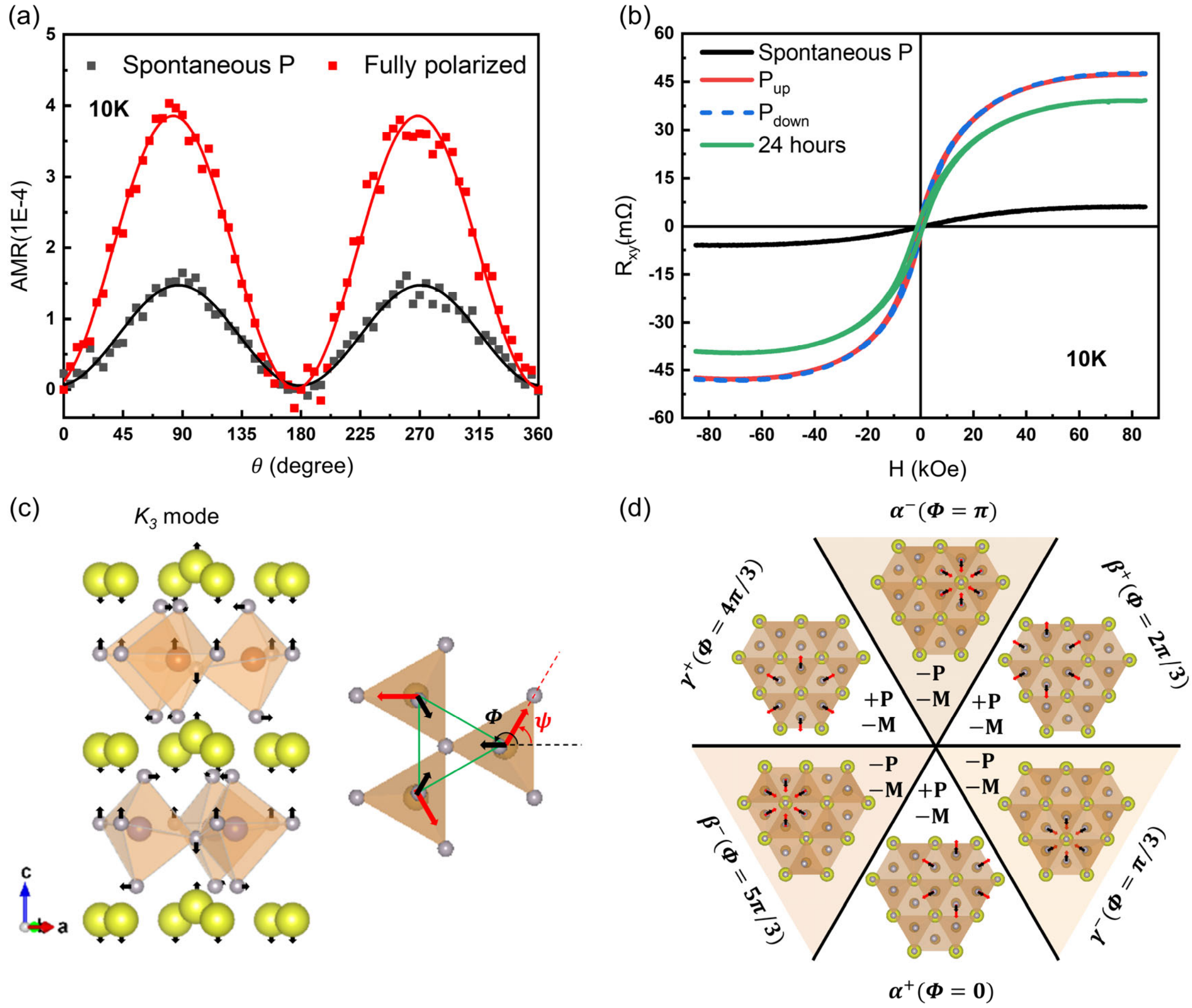

**Fig. 4.** (**a**) AMR measurements at 10K on the 1.5 UC *h*-$LuFeO_3$ after applying a strong electric field of 1.5 V/100 nm across parallel metal electrodes. The black curve represents the initial spontaneous-polarization state, while the red curve corresponds to the magnetic signal after applying the electric field, with the ferroelectric polarization is aligned in a single direction. (**b**) AHE signal under different ferroelectric polarization conditions at 10 K. The black solid line represents the initial spontaneous-polarization state, while the red dashed line shows the AHE signal after full positive polarization (+Z). The blue dashed line corresponds to the AHE signal when the sample is fully polarized in the reverse direction (-Z). The green solid line corresponds to the relaxed multidomain state obtained after the poled sample was left undisturbed for 24 h. (**c**) Schematic illustration of the $K_3$ lattice distortion and the definition of phase Φ and spin orientation $\psi$. The black arrows denote the structural trimerization associated with the $FeO_5$ bipyramids, and the red arrows indicate the spin orientation. Within a single ferroelectric domain, Φ and $\psi$ are locked with a phase difference of $n\pi$. (**d**) Schematic of the six degenerate ferroelectric domain states defined by the phase Φ of the $K_3$ distortion. In a saturated single-domain state, the weak net moment is maximized. Across a ferroelectric domain wall, the local phase relation between Φ and $\psi$ is disrupted, giving rise to clamped antiferromagnetic walls and a reduced net magnetic response in the multidomain state.

**Supplementary Materials for**

# Multiferroicity in the two-dimensional limit in hexagonal $LuFeO_3$ films

Huilin Lai[1,2], Junyu Tan[1,2], Jinfeng Zhai[1,2], Yang Shi[1,2], Lili Feng[1,2], Huanyu Zhang[1,2], Chuanrui Huo[3], Chuhang Liu[4], Lijun Wu[4], Lifeng Yin[1,2,5,6,7], Hangwen Guo[1,2,5], Jun Chen[3], Xiaoshan Xu[8], Jun Zhao[1,2,5,6], Yimei Zhu[4], Shiqing Deng[3*], Wenbin Wang[1,2,5*] and Jian Shen[1,2,5,6,7*]

[1]State Key Laboratory of Surface Physics and Institute for Nanoelectronic Devices and Quantum Computing, Fudan University, Shanghai 200433, China.

[2]Department of Physics, Fudan University, Shanghai 200433, China.

[3]Beijing Advanced Innovation Center for Materials Genome Engineering, University of Science and Technology Beijing, Beijing, 100083, P.R. China.

[4]Condensed Matter Physics and Materials Science Department, Brookhaven National Laboratory, Upton, NY 11973, USA.

[5]Hefei National Laboratory, Hefei 230088, China.

[6]Shanghai Research Center for Quantum Sciences, Shanghai 201315, China.

[7]Zhangjiang Fudan International Innovation Center, Fudan University, Shanghai 201210, China.

[8]Department of Physics and Astronomy, University of Nebraska, Lincoln, Nebraska 68588, USA.

*Corresponding author. Email: sqdeng@ustb.edu.cn; wangwb@fudan.edu.cn; shenj5494@fudan.edu.cn

## Methods

### Sample Deposition and Device Preparation

All samples were grown via pulsed laser deposition (PLD) on yttrium-stabilized zirconium oxide (YSZ) (111) substrates. Prior to growth, the base pressure in the main chamber was maintained below $1\times10^{-8}$ Torr, with an oxygen pressure of 20 mTorr during deposition. Heating was achieved using an infrared laser, which uniformly illuminated a 10 mm × 10 mm SiC absorber. The 5 mm × 5 mm YSZ substrate was placed directly on the SiC, with the substrate temperature held at 780°C and monitored in real-time via infrared pyrometry. Growth was conducted with a KrF excimer laser at a 3 Hz repetition rate and energy density of ~1.5 J/cm$^2$. To ensure uniform deposition, the target stage was rotated continuously, and laser scanning was applied across the target surface.

Reflection high-energy electron diffraction (RHEED) was employed for in situ monitoring, confirming atomic-level smoothness and precise layer-by-layer thickness control through periodic intensity oscillations. Initially, a 3-layer $Lu_2O_3$ buffer layer was deposited on YSZ (111) substrate under identical growth conditions as *h*-$LuFeO_3$. The $Lu_2O_3$ buffer layer enhances lattice matching and stabilizes the multiferroic properties of *h*-$LuFeO_3$, ensuring optimal structural compatibility. Notably, $Lu_2O_3$ itself is neither magnetic nor ferroelectric, serving solely as a structural buffer to support the growth of high-quality multiferroic *h*-$LuFeO_3$ films.

For ferroelectric measurements, *h*-$LuFeO_3$/$Lu_2O_3$/ITO/YSZ multilayer samples were prepared. A 40 nm thick $In_2O_3$:$SnO_2$ (ITO) layer, acting as the bottom electrode, was grown under the same conditions as the *h*-$LuFeO_3$ layer.

For magnetic measurements, Pt/*h*-$LuFeO_3$/$Lu_2O_3$/YSZ multilayer samples were fabricated. After growing *h*-$LuFeO_3$, the temperature was lowered to ambient, the oxygen flow was stopped, and the chamber pressure was set to $1\times10^{-8}$ Torr. Pt was then in situ deposited at a laser energy density of 2 J/cm$^2$, resulting in a 4.5 nm Pt film. Post-growth structural analysis confirmed the Pt layer was well-crystallized with a (111) orientation. Hall bar devices for AMR and AHE measurements were fabricated using lithography. To isolate the magnetic contribution from *h*-$LuFeO_3$, control devices with a 2 nm Cu interlayer—deposited under identical conditions to Pt—were also fabricated. This control helps to confirm that the observed magnetic properties are due to *h*-$LuFeO_3$ by comparing the effects of Cu and Pt interlayers.

### Structural Characterization

Throughout the growth process, RHEED patterns were used for in situ monitoring of both structure and thickness. The RHEED patterns confirmed layer-by-layer growth for all samples, with atomic-scale smoothness and no evidence of extraneous phases. Each RHEED oscillation period corresponded to half a unit cell of *h*-$LuFeO_3$, allowing precise control over thickness, which was rigorously monitored for all samples.

After growth, detailed structural analysis was performed using X-ray diffraction (XRD) with a Bruker D8 Discover. The fitted XRR curves confirmed values consistent with RHEED calibrations. XRD analysis showed high crystallinity, with no detectable impurity phases. For thicker samples, rocking curve measurements revealed a full width at half maximum (FWHM) of approximately $0.13^{\circ}$, indicating excellent crystal qualit.

For devices fabricated to measure ferroelectric and magnetic properties, XRD characterization confirmed the presence of ITO (111) and Pt (111) peaks in the *h*-$LuFeO_3$/$Lu_2O_3$/ITO/YSZ and Pt/*h*-$LuFeO_3$/$Lu_2O_3$/YSZ samples.

**Microscopic characterization:**

Atomic-scale characterizations were conducted using a JEOL ARM 200CF aberration-corrected STEM operated at 200 kV. The microscope is equipped with a cold-field emission gun, double correctors, and a Gatan GIF Quantum ER Energy Filter with DualEELS, and a K3 camera. A convergent semi-angle of 21.2 mrad was used for both STEM imaging and EELS mapping. The inner collection angles for STEM-HAADF imaging and EELS mapping were 67 and 88 mrad, respectively. EELS mapping was carried out with a dispersion of 0.1 eV/channel and a dwell time of 0.1 μs/pixel.

**Ferroelectric Properties Measurements**

Ferroelectric properties of the 1.5 UC *h*-$LuFeO_3$ were characterized using RHEED, STEM, and Piezoresponse Force Microscopy (PFM). RHEED patterns indicated structural changes associated with the paraelectric-to-ferroelectric transition, marked by additional streaks at 1/3 and 2/3 positions, corresponding to in-plane unit cell tripling ($P6_3cm$ phase) relative to the paraelectric phase ($P6_3/mmc$). The presence of these streaks confirms a ferroelectric structure in 1.5 UC *h*-$LuFeO_3$. STEM imaging revealed tilting of $FeO_5$ bipyramids and corrugation of Lu layers, consistent with unit cell tripling and confirming that the material adopts a ferroelectric structure at room temperature.

To directly observe polarization switching, we introduced a 40 nm $In_2O_3$:$SnO_2$(ITO) bottom electrode between the YSZ (111) substrate and the h-$LuFeO_3$ layers, PFM measurements were performed on Bruker Dimension Icon equipped with SCM-PIT-V2 Pt/Ir-coated silicon cantilevers. Additionally, we investigated the retention properties of the ferroelectric material. A voltage sequence of -2 V, +2 V, and -2 V was applied over a 12×12 μm area. Morphology characterization confirmed that the surface remained unchanged during this process. Immediately afterward, PFM phase and amplitude images were captured while maintaining the probe's position. Subsequently, the same area was scanned at intervals of 1 hour, 2 hours, and 3.5 hours, recording the phase and amplitude images. To quantitatively analyze the retention properties, we extracted the average phase values of a 2×2 μm region at the center of the negatively polarized area and plotted their evolution over time. Fitting these data confirmed a slow decay, consistent with robust retention properties.

**Magnetic Properties Measurements**

The magnetic properties of the samples were evaluated using anisotropic magnetoresistance (AMR) and anomalous Hall effect (AHE) measurements. Samples with $h$-$LuFeO_3$ layers of varying thicknesses—22 UC (25.7 nm), 9 UC (10.5 nm), 4 UC (4.7 nm), and 1.5 UC (1.8 nm)—were in situ capped with a 4.5 nm Pt layer and patterned into 600 μm × 100 μm Hall bars. During measurements, a 100 μA current was applied along the x-axis. Measurements were conducted using a Quantum Design PPMS-9 with EverCool II, equipped with a rotational sample rod, enabling variation of the relative orientation between the sample, magnetic field, and current during the experiments. Samples were initially zero-field-cooled to low temperature, then subjected to an out-of-plane magnetic field ramping from 0 to 8.5 T. Samples were rotated within the *x*-*z* plane, followed by field removal and warming to the next temperature setpoint. For thicker samples, SQUID magnetometry was also conducted using a Quantum Design SQUID-VSM, where zero-field-cooled (ZFC) and field-cooled (FC) magnetization curves were recorded in the out-of-plane direction. The signal intensity of anisotropic magnetoresistance (AMR) is calculated using the formula : $AMR = [R(\theta) - R(\theta = 0)]/R(\theta = 0)$. The magnetoelectric coupling measurement was conducted by applying a ferroelectric switching voltage to the device using an external setup. A 10 μm mica layer, affixed to the device surface with resin, served as the dielectric layer. The entire device was then placed inside the high-voltage ceramic chamber of the ferroelectric measurement system, specifically the MODEL 609E-6 High Voltage Amplifier. Within the chamber, a pair of parallel copper plates formed a parallel-plate capacitor. By applying a voltage across the copper electrodes, the ferroelectric polarization of the device was switched.

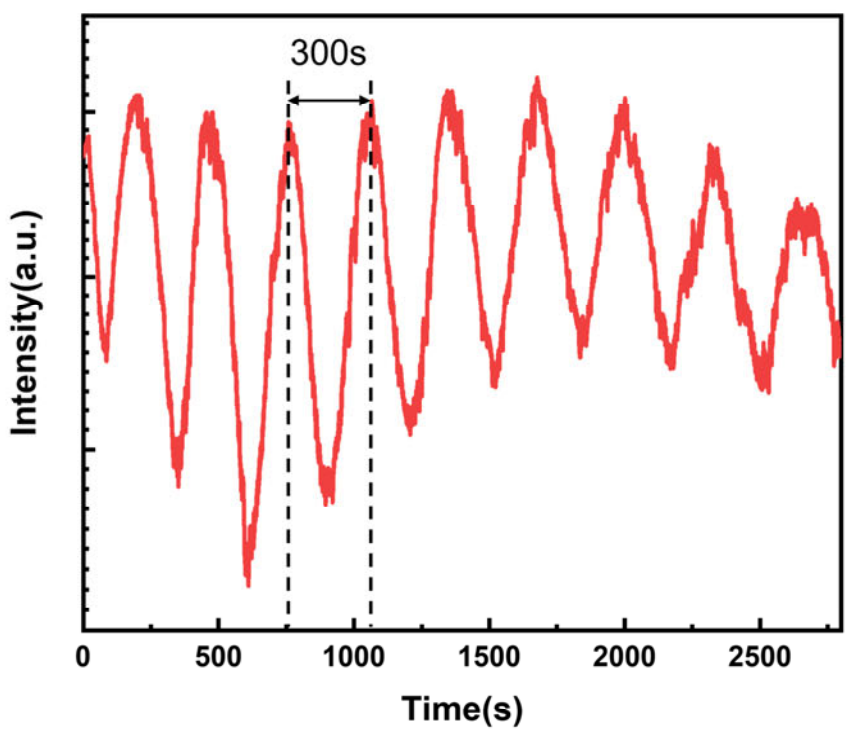

**Fig. S1** Real-time reflection high-energy electron diffraction (RHEED) during the growth of *h*-$LuFeO_3$ films showed that each RHEED oscillation corresponded to the completion of half a unit cell of *h*-$LuFeO_3$, allowing precise measurement of the film's thickness during deposition. The periodic intensity changes confirmed layer-by-layer growth.

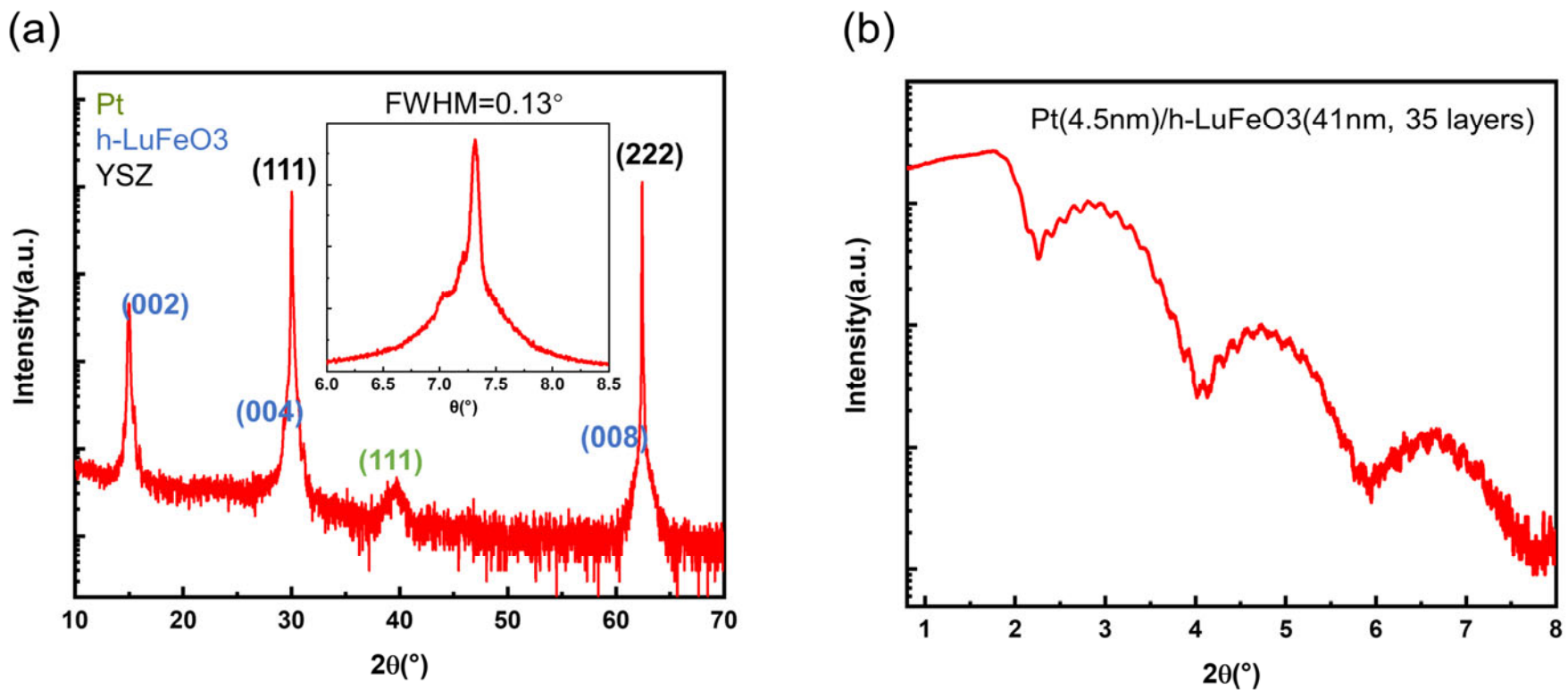

**Fig. S2** Crystallinity and thickness characterization of *h*-$LuFeO_3$ films. (**a**) The XRD pattern shows sharp and intense (002), (004), and (008) peaks for the *h*-$LuFeO_3$ film (blue annotations), along with (111) peaks for the Pt layer (green annotations) and the YSZ substrate (black annotations). This indicates that the films are oriented along the (001) direction with no detectable impurity phases. The full width at half maximum (FWHM) of the *h*-$LuFeO_3$ (002) peak is approximately 0.13°, suggesting high film quality. (**b**) XRR curves. The smaller oscillation period corresponds to the thickness of *h*-$LuFeO_3$, while the larger oscillation period corresponds to the Pt layer thickness. Fitting results show that the Pt layer is 4.5 nm thick, and the *h*-$LuFeO_3$ film consists of 35 unit cells (UC) with a total thickness of 41 nm.

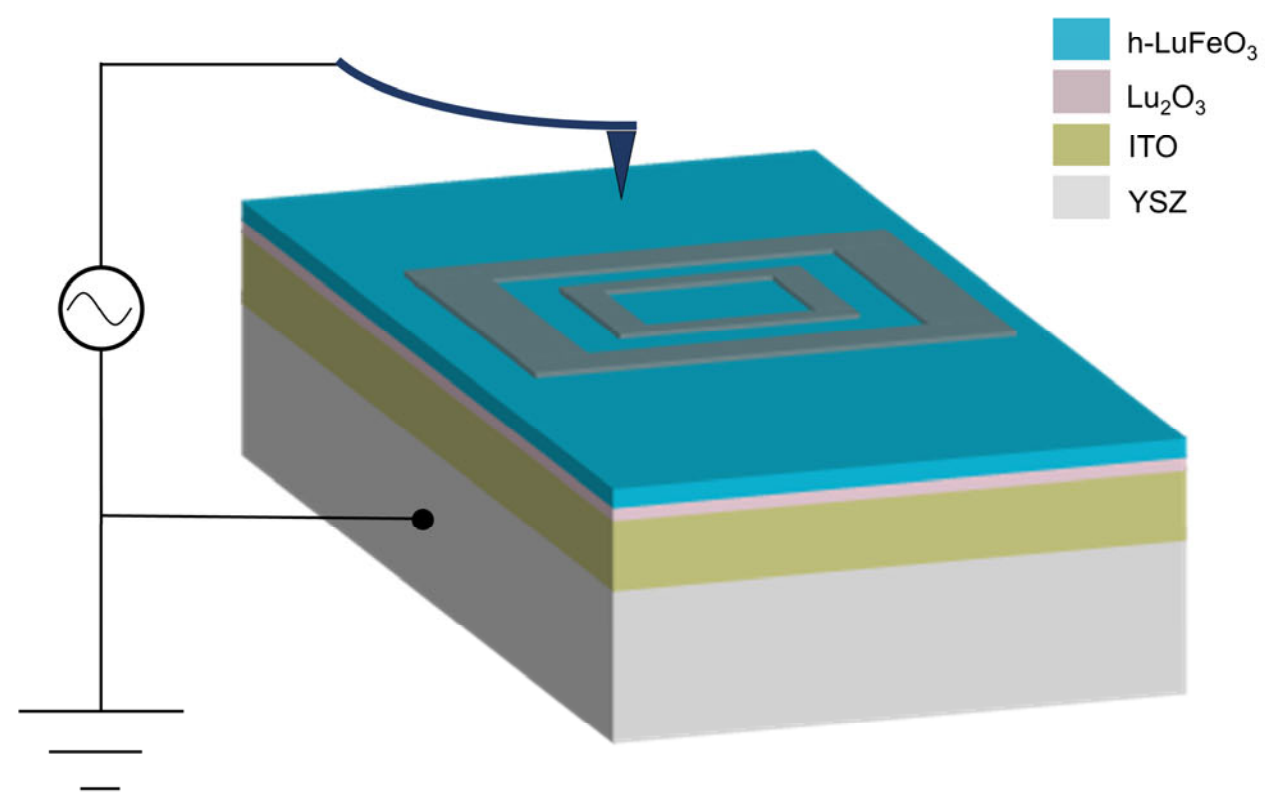

**Fig. S3:** Schematic of the *h*-$LuFeO_3$ based device for polarization switching measurements. The device includes of a 40 nm $In_2O_3$:$SnO_2$ (ITO) bottom electrode positioned between a YSZ (111) substrate and the *h*-$LuFeO_3$/$Lu_2O_3$ layers. This configuration allows the measurement of polarization switching behavior using PFM, where the polarization of *h*-$LuFeO_3$ is reversed by applying a voltage to the tip.

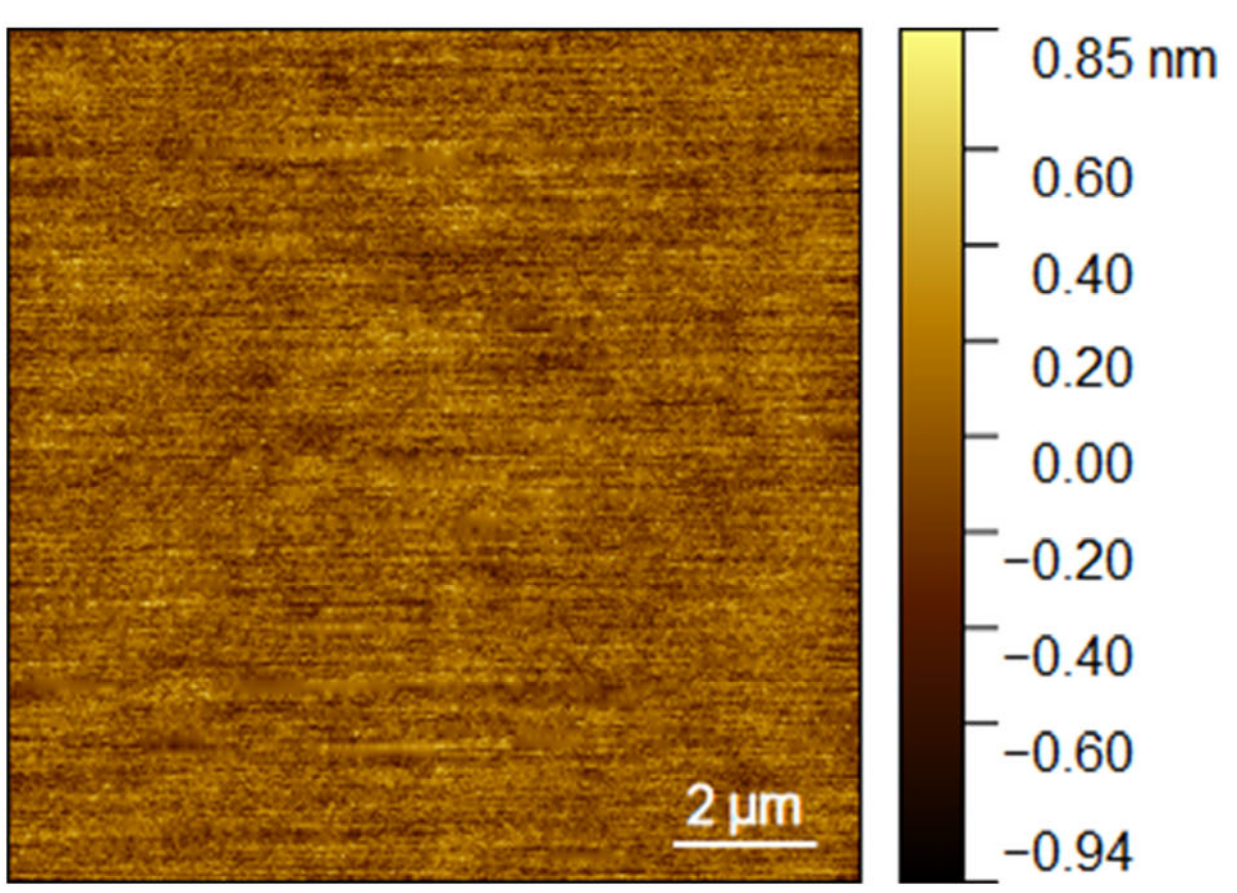

**Fig. S4:** Surface morphology of *h*-$LuFeO_3$ after polarization switching via PFM. The image shows the surface topography of the sample following the application of an electric field to write ferroelectric domains. No significant structural changes or surface roughness (RMS=152.2pm) were observed, indicating that the applied electric field did not cause notable morphological alterations. The uniformity of the surface confirms the stability of the material's structure after polarization switching.

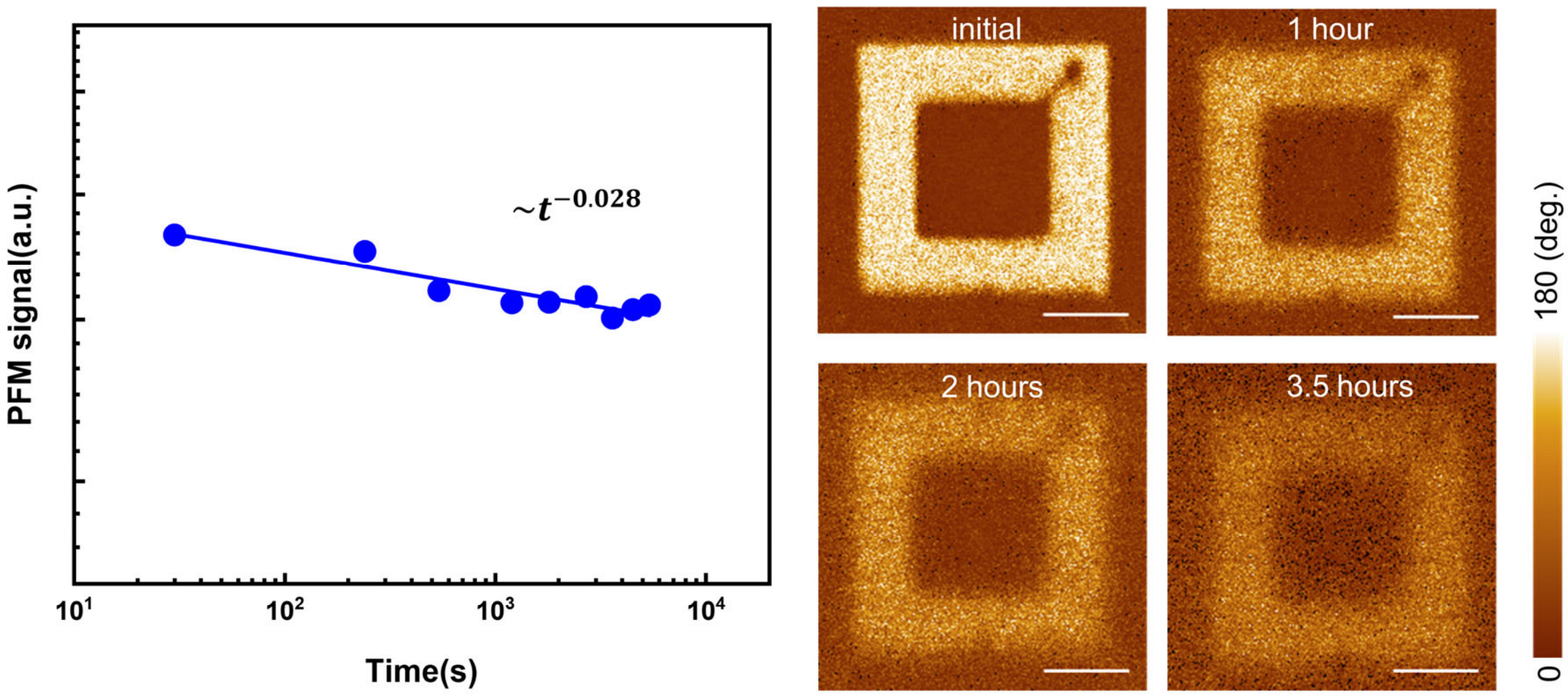

**Fig. S5**: PFM signal as a function of delay time measured at room temperature after polarization switching in a 1.5 UC *h*-$LuFeO_3$ film. The solid line represents the simulated curve of signal decay, which follows a power-law decay $P(t) \propto t^{-\alpha}$, where t is the decay time and a is the decay exponent. The decay exponent $\alpha$ of 0.028 for the 1.5 UC *h*-$LuFeO_3$ film indicates an exceptionally slow relaxation, underscoring the remarkable stability of polarization at this ultrathin limit. The images on the right show the initial phase image and those taken at 1, 2, and 3.5 hours after polarization.

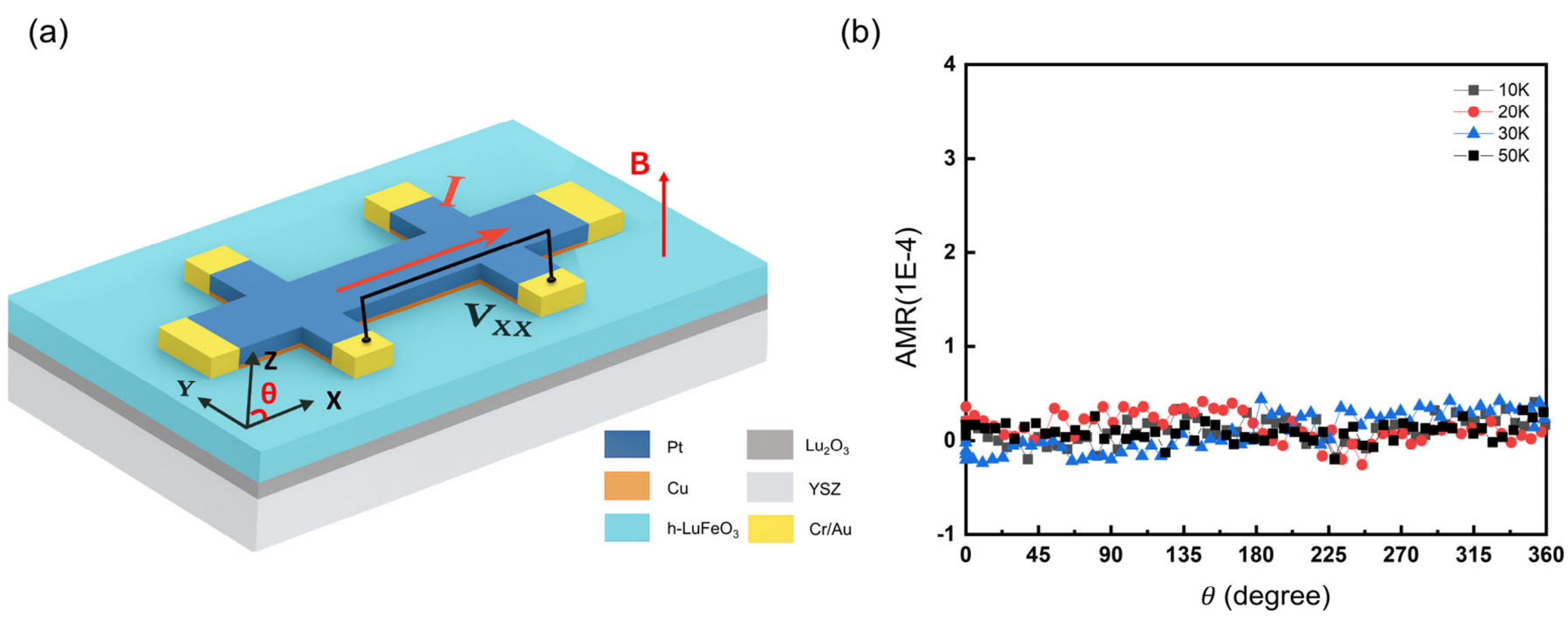

**Fig. S6:** Effect of a 2 nm Cu spacer on the anisotropic magnetoresistance (AMR) signals in *h*-$LuFeO_3$. (**a**) Device schematic: To eliminate potential proximity effects, A 2 nm Cu spacer was introduced between the Pt and the 22 UC *h*-$LuFeO_3$ layers to eliminate potential proximity effects. (**b**) At various temperatures, no AMR signals were detected, confirming that the observed magnetic properties are intrinsic to *h*-$LuFeO_3$ and not influenced by interface effects with the Pt electrodes. This experiment demonstrates the inherent magnetism of the *h*-$LuFeO_3$ layer.

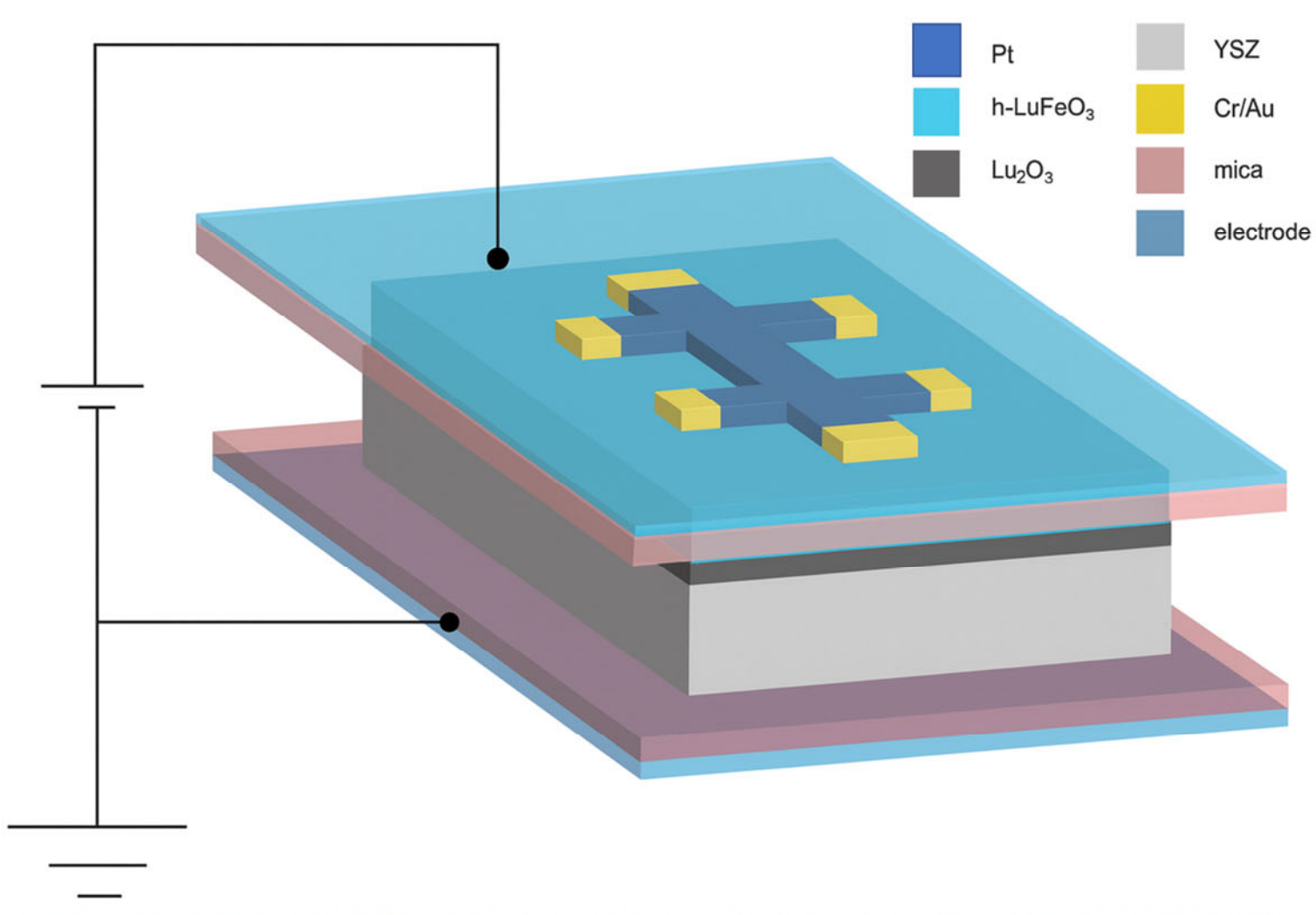

**Fig. S7:** Device schematic: A 10 μm mica layer was attached to the front side of the device using resin as the dielectric layer. The entire device was then placed inside the high-voltage ceramic chamber of the ferroelectric tester, which contains a pair of parallel copper plates forming a parallel-plate capacitor. Voltage was applied to the copper electrodes to switch the ferroelectric polarization of the device. An electric field of 1.5 V/100 nm was applied, and measurements were conducted at 10 K.

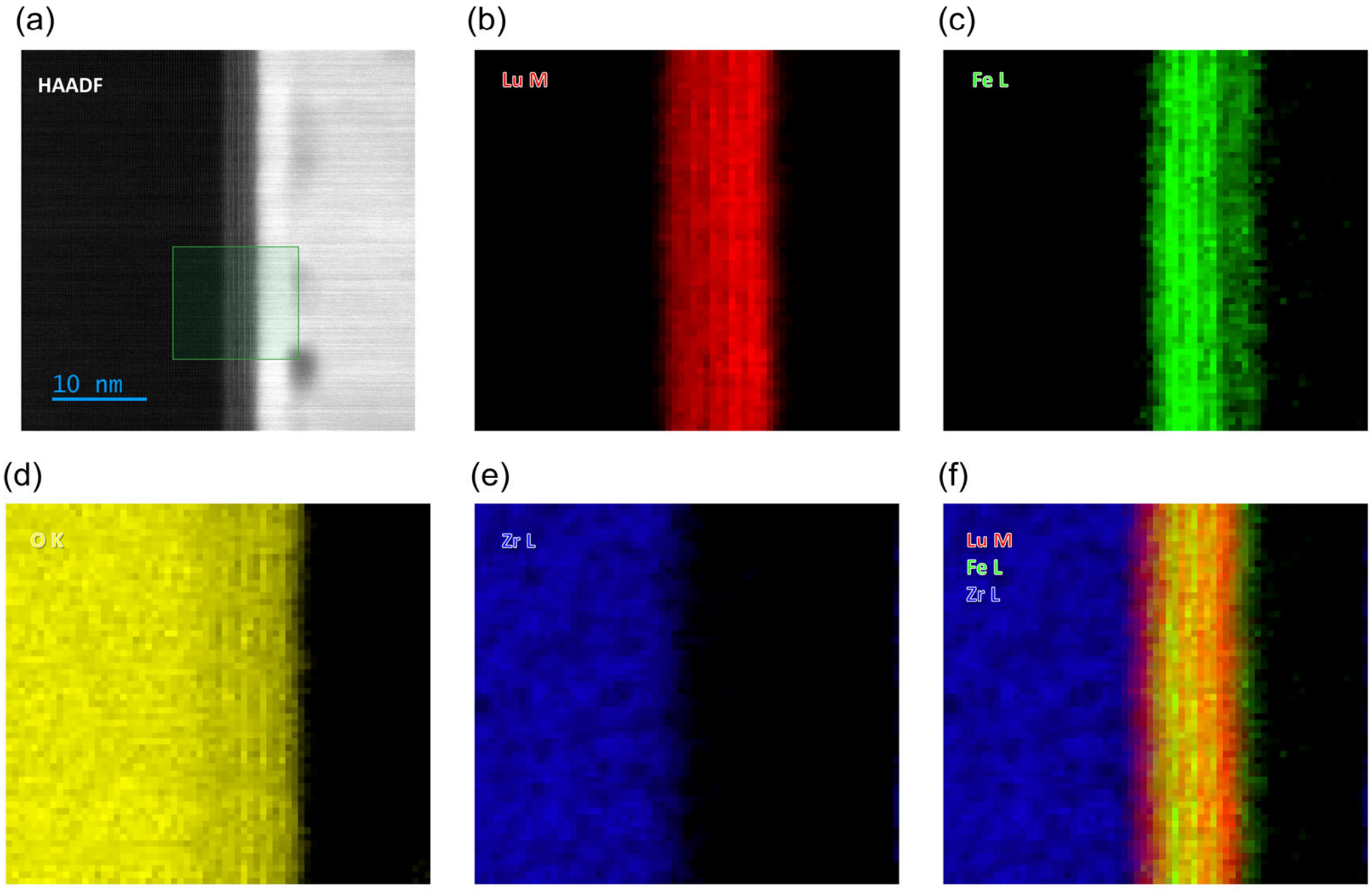

**Fig. S8:** EELS elemental maps of Lu, Fe, and O over an extended region, showing uniform spatial distributions across the film. The Lu M-edge and Fe L-edge signals clearly resolve the atomic layers, confirming the high compositional uniformity of the film.

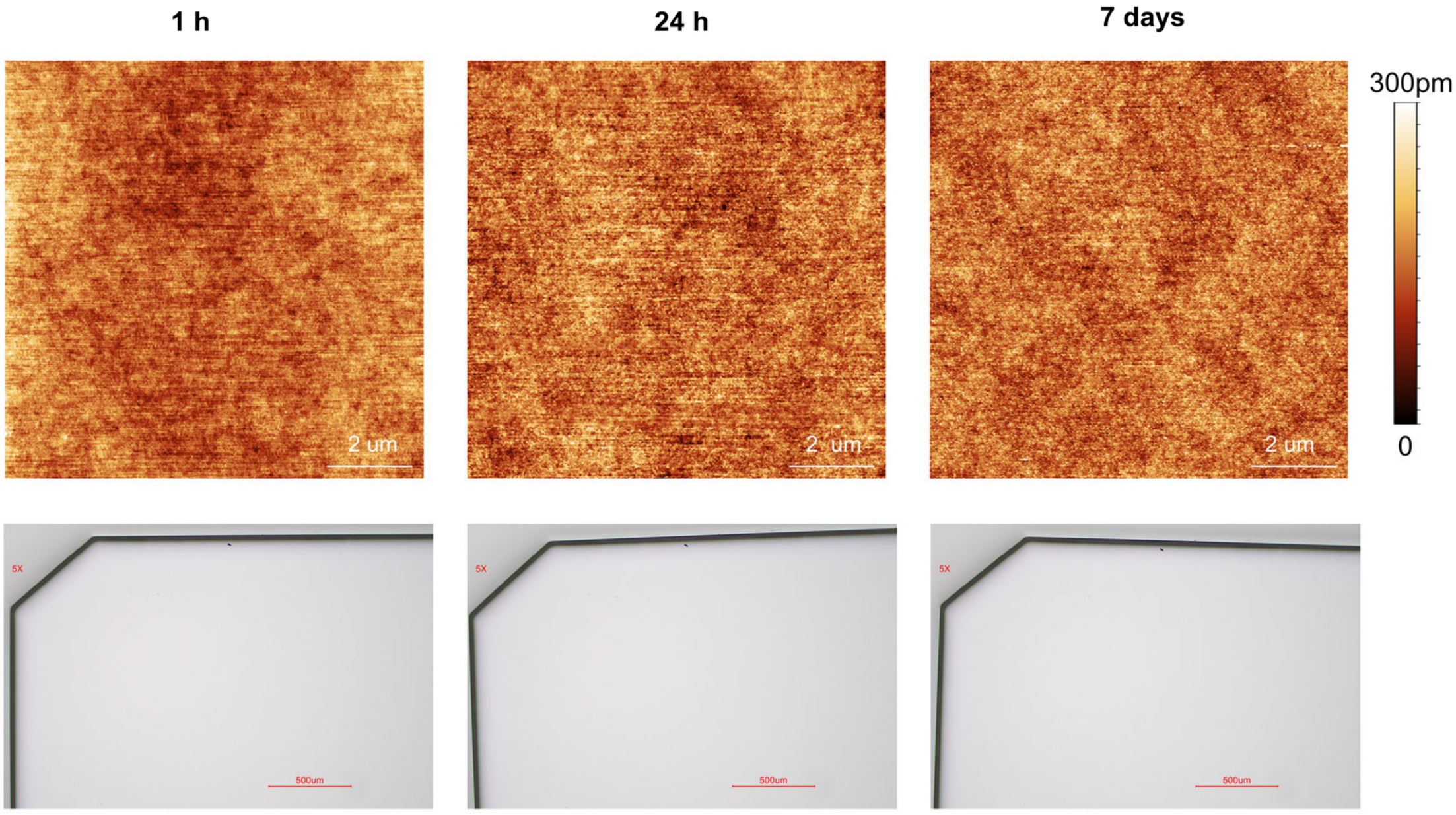

**Fig. S9:** Atomic force microscopy (top) and optical microscopy (bottom) were conducted on 1.5-unit-cell *h*-$LuFeO_3$ films at different intervals after growth (1 h, 24 h, and 1 week). All films were stored under ambient conditions in a standard dry cabinet without encapsulation. No detectable degradation or morphological changes were observed, demonstrating excellent air stability even at the 1.5-unit-cell limit.